# 1. DARK ENERGY

Written November 2013 by M. J. Mortonson (UCB, LBL), D. H. Weinberg (OSU), and M. White (UCB, LBL).

## 1.1. Repulsive Gravity and Cosmic Acceleration

In the first modern cosmological model, Einstein [1] modified his field equation of General Relativity (GR), introducing a "cosmological term" that enabled a solution with time-independent, spatially homogeneous matter density $\rho_m$ and constant positive space curvature. Although Einstein did not frame it this way, one can view the "cosmological constant" $\Lambda$ as representing a constant energy density of the vacuum [2], whose repulsive gravitational effect balances the attractive gravity of matter and thereby allows a static solution. After the development of dynamic cosmological models [3,4] and the discovery of cosmic expansion [5], the cosmological term appeared unnecessary, and Einstein and de Sitter [6] advocated adopting an expanding, homogeneous and isotropic, spatially flat, matter-dominated universe as the default cosmology until observations dictated otherwise. Such a model has matter density equal to the critical density, $\Omega_m \equiv \rho_m/\rho_c = 1$, and negligible contribution from other energy components [7].

By the mid-1990s, Big Bang cosmology was convincingly established, but the Einstein-de Sitter model was showing numerous cracks, under the combined onslaught of data from the cosmic microwave background (CMB), large scale galaxy clustering, and direct estimates of the matter density, the expansion rate ($H_0$), and the age of the Universe. Introducing a cosmological constant offered a potential resolution of many of these tensions. In the late 1990s, supernova surveys by two independent teams provided direct evidence for accelerating cosmic expansion [8,9], establishing the cosmological constant model (with $\Omega_m \approx 0.3$, $\Omega_\Lambda \approx 0.7$) as the preferred alternative to the $\Omega_m = 1$ scenario. Shortly thereafter, CMB evidence for a spatially flat universe [10,11], and thus for $\Omega_{tot} \approx 1$, cemented the case for cosmic acceleration by firmly eliminating the free-expansion alternative with $\Omega_m \ll 1$ and $\Omega_\Lambda = 0$. Today, the accelerating universe is well established by multiple lines of independent evidence from a tight web of precise cosmological measurements.

As discussed in the Big Bang Cosmology article of this *Review* (Sec. 19), the scale factor $R(t)$ of a homogeneous and isotropic universe governed by GR grows at an accelerating rate if the pressure $p < -\frac{1}{3}\rho$. A cosmological constant has $\rho_\Lambda = $ const. and pressure $p_\Lambda = -\rho_\Lambda$ (see Eq. 19.10), so it will drive acceleration if it dominates the total energy density. However, acceleration could arise from a more general form of "dark energy" that has negative pressure, typically specified in terms of the equation-of-state-parameter $w = p/\rho$ ($= -1$ for a cosmological constant). Furthermore, the conclusion that acceleration requires a new energy component beyond matter and radiation relies on the assumption that GR is the correct description of gravity on cosmological scales. The title of this article follows the common but inexact usage of "dark energy" as a catch-all term for the origin of cosmic acceleration, regardless of whether it arises from a new form of energy or a modification of GR. Our account here draws on the much longer review of cosmic acceleration by Ref. [12], which provides background explanation and extensive literature references for most of the points in this article, but is less up to date in its description of current empirical constraints.

Chapter 25 from the Particle Data Group Review of Particle Physics, 2013 partial update for the 2014 division. Other chapters referred to in this review can be found at http://pdg.lbl.gov.

Below we will use the abbreviation ΛCDM to refer to a model with cold dark matter, a cosmological constant, inflationary initial conditions, and standard radiation and neutrino content. We will use "flat ΛCDM" to further specify a flat universe with $\Omega_{\text{tot}} = 1$. We will use wCDM to denote a model with the same assumptions (including flatness) but a free, constant value of $w$.

## 1.2.  Theories of Cosmic Acceleration

### 1.2.1.  *Dark Energy or Modified Gravity?*:

A cosmological constant is the mathematically simplest, and perhaps the physically simplest, theoretical explanation for the accelerating universe. The problem is explaining its unnaturally small magnitude, as discussed in Sec. 19.4.7 of this *Review*. An alternative (which still requires finding a way to make the cosmological constant zero or at least negligibly small) is that the accelerating cosmic expansion is driven by a new form of energy such as a scalar field [13] with potential $V(\phi)$. The energy density and pressure of the field $\phi(\mathbf{x})$ take the same forms as for inflationary scalar fields, given in Eq. (19.52) of the Big Bang Cosmology article. In the limit that $\frac{1}{2}\dot{\phi}^2 \ll |V(\phi)|$, the scalar field acts like a cosmological constant, with $p_\phi \approx -\rho_\phi$. In this scenario, today's cosmic acceleration is closely akin to the epoch of inflation, but with radically different energy and timescale.

More generally, the value of $w = p_\phi/\rho_\phi$ in scalar field models evolves with time in a way that depends on $V(\phi)$ and on the initial conditions $(\phi_i, \dot{\phi}_i)$; some forms of $V(\phi)$ have attractor solutions in which the late-time behavior is insensitive to initial values. Many forms of time evolution are possible, including ones where $w$ is approximately constant and broad classes where $w$ "freezes" towards or "thaws" away from $w = -1$, with the transition occurring when the field comes to dominate the total energy budget. If $\rho_\phi$ is even approximately constant, then it becomes dynamically insignificant at high redshift, because the matter density scales as $\rho_{\text{m}} \propto (1+z)^3$. "Early dark energy" models are ones in which $\rho_\phi$ is a small but not negligible fraction (*e.g.*, a few percent) of the total energy throughout the matter and radiation dominated eras, tracking the dominant component before itself coming to dominate at low redshift.

Instead of introducing a new energy component, one can attempt to modify gravity in a way that leads to accelerated expansion [14]. One option is to replace the Ricci scalar $\mathcal{R}$ with a function $\mathcal{R} + f(\mathcal{R})$ in the gravitational action [15]. Other changes can be more radical, such as introducing extra dimensions and allowing gravitons to "leak" off the brane that represents the observable universe (the "DGP" model [16]) . The DGP example has inspired a more general class of "galileon" and massive gravity models. Constructing viable modified gravity models is challenging, in part because it is easy to introduce theoretical inconsistencies (such as "ghost" fields with negative kinetic energy) but above all because GR is a theory with many high-precision empirical successes on solar system scales [17]. Modified gravity models typically invoke screening mechanisms that force model predictions to approach those of GR in regions of high density or strong gravitational potential. Screening offers potentially distinctive signatures, as the strength of gravity (*i.e.*, the effective value of $G_{\text{N}}$) can vary by order unity in environments with different gravitational potentials.

More generally, one can search for signatures of modified gravity by comparing the history of cosmic structure growth to the history of cosmic expansion. Within GR, these two are linked by a consistency relation, as described below (Eq. (1.2)). Modifying gravity can change the predicted rate of structure growth, and it can make the growth rate dependent on scale or environment. In some circumstances, modifying gravity alters the combinations of potentials responsible for gravitational lensing and the dynamics of non-relativistic tracers (such as galaxies or stars) in different ways (see Sec. 19.4.7 in this *Review*), leading to order unity mismatches between the masses of objects inferred from lensing and those inferred from dynamics in unscreened environments.

At present there are no fully realized and empirically viable modified gravity theories that explain the observed level of cosmic acceleration. The constraints on $f(\mathcal{R})$ models now force them so close to GR that they cannot produce acceleration without introducing a separate dark energy component [18]. The DGP model is empirically ruled out by several tests, including the expansion history, the integrated Sachs-Wolfe effect, and redshift-space distortion measurements of the structure growth rate [19]. The elimination of these models should be considered an important success of the program to empirically test theories of cosmic acceleration. However, it is worth recalling that there was no fully realized gravitational explanation for the precession of Mercury's orbit prior to the completion of GR in 1915, and the fact that no complete and viable modified gravity theory exists today does not mean that one will not arise in the future. In the meantime, we can continue empirical investigations that can tighten restrictions on such theories or perhaps point towards the gravitational sector as the origin of accelerating expansion.

### 1.2.2. *Expansion History and Growth of Structure*:

The main line of empirical attack on dark energy is to measure the history of cosmic expansion and the history of matter clustering with the greatest achievable precision over a wide range of redshift. Within GR, the expansion rate $H(z)$ is governed by the Friedmann equation (see the articles on Big Bang Cosmology and Cosmological Parameters—Secs. 19 and 21 in this *Review*). For dark energy with an equation of state $w(z)$, the cosmological constant contribution to the expansion, $\Omega_\Lambda$, is replaced by a redshift-dependent contribution with the evolution of the dark energy density following from Eq. (19.10),

$$\Omega_{\mathrm{DE}} \frac{\rho_{\mathrm{DE}}(z)}{\rho_{\mathrm{DE}}(z=0)} = \Omega_{\mathrm{DE}} \exp\left[3\int_0^z \left[1 + w(z')\right] \frac{dz'}{1+z'}\right] \;\; = \;\; \Omega_{\mathrm{DE}}(1+z)^{3(1+w)}, \quad (1.1)$$

where the second equality holds for constant $w$. If $\Omega_{\mathrm{m}}$, $\Omega_{\mathrm{r}}$, and the present value of $\Omega_{\mathrm{tot}}$ are known, then measuring $H(z)$ pins down $w(z)$. (Note that $\Omega_{\mathrm{DE}}$ is the same quantity denoted $\Omega_{\mathrm{v}}$ in Sec. 19, but we have adopted the DE subscript to avoid implying that dark energy is necessarily a vacuum effect.)

While some observations can probe $H(z)$ directly, others measure the distance-redshift relation. The basic relations between angular diameter distance or luminosity distance and $H(z)$ are given in Ch. 19 —and these are generally unaltered in time-dependent dark energy or modified gravity models. For convenience, in later sections, we will sometimes refer to the comoving angular distance, $D_{\mathrm{A,c}}(z) = (1+z)D_{\mathrm{A}}(z)$.

In GR-based linear perturbation theory, the density contrast $\delta(\mathbf{x}, t) \equiv \rho(\mathbf{x}, t)/\bar{\rho}(t) - 1$ of pressureless matter grows in proportion to the linear growth function $G(t)$ (not to be confused with the gravitational constant $G_N$), which follows the differential equation

$$\ddot{G} + 2H(z)\dot{G} - \frac{3}{2}\Omega_m H_0^2 (1+z)^3 G = 0 \; . \tag{1.2}$$

To a good approximation, the logarithmic derivative of $G(z)$ is

$$f(z) \equiv -\frac{d\ln G}{d\ln(1+z)} \approx \left[\Omega_m (1+z)^3 \frac{H_0^2}{H^2(z)}\right]^\gamma \; , \tag{1.3}$$

where $\gamma \approx 0.55$ for relevant values of cosmological parameters [20]. In an $\Omega_m = 1$ universe, $G(z) \propto (1+z)^{-1}$, but growth slows when $\Omega_m$ drops significantly below unity. One can integrate Eq. (1.3) to get an approximate integral relation between $G(z)$ and $H(z)$, but the full (numerical) solution to Eq. (1.2) should be used for precision calculations. Even in the non-linear regime, the amplitude of clustering is determined mainly by $G(z)$, so observations of non-linear structure can be used to infer the linear $G(z)$, provided one has good theoretical modeling to relate the two.

In modified gravity models the growth rate of gravitational clustering may differ from the GR prediction. A general strategy to test modified gravity, therefore, is to measure both the expansion history and the growth history to see whether they yield consistent results for $H(z)$ or $w(z)$.

### 1.2.3. *Parameters*:

Constraining a general history of $w(z)$ is nearly impossible, because the dark energy density, which affects $H(z)$, is given by an integral over $w(z)$, and distances and the growth factor involve a further integration over functions of $H(z)$. Oscillations in $w(z)$ over a range $\Delta z/(1+z) \ll 1$ are therefore extremely difficult to constrain. It has become conventional to phrase constraints or projected constraints on $w(z)$ in terms of a linear evolution model,

$$w(a) = w_0 + w_a(1-a) = w_p + w_a(a_p - a) \; , \tag{1.4}$$

where $a \equiv (1+z)^{-1}$, $w_0$ is the value of $w$ at $z = 0$, and $w_p$ is the value of $w$ at a "pivot" redshift $z_p \equiv a_p^{-1} - 1$, where it is best constrained by a given set of experiments. For typical data combinations, $z_p \approx 0.5$. This simple parameterization can provide a good approximation to the predictions of many physically motivated models for observables measured with percent-level precision. A widely used "Figure of Merit" (FoM) for dark energy experiments [21] is the projected combination of errors $[\sigma(w_p)\sigma(w_a)]^{-1}$. Ambitious future experiments with 0.1–0.3% precision on observables can constrain richer descriptions of $w(z)$, which can be characterized by principal components.

There has been less convergence on a standard parameterization for describing modified gravity theories. Deviations from the GR-predicted growth rate can be described by a deviation $\Delta\gamma$ in the index of Eq. (1.3), together with an overall multiplicative offset relative to the $G(z)$ expected from extrapolating the CMB-measured fluctuation amplitude to low redshift. However, these two parameters may not accurately capture the growth predictions of all physically interesting models. Another important parameter to constrain is the ratio of the gravitational potentials governing space curvature and the acceleration of non-relativistic test particles. The possible phenomenology of modified gravity models is rich, which enables many consistency tests but complicates the task of constructing parameterized descriptions.

The more general set of cosmological parameters is discussed elsewhere in this *Review* (Sec. 21), but here we highlight a few that are particularly important to the dark energy discussion:

- The dimensionless Hubble parameter $h \equiv H_0/100 \, \mathrm{km \, s^{-1} \, Mpc^{-1}}$ determines the present day value of the critical density and the overall scaling of distances inferred from redshifts.

- $\Omega_\mathrm{m}$ and $\Omega_\mathrm{tot}$ affect the expansion history and the distance-redshift relation.

- The sound horizon $r_\mathrm{s} = \int_0^{t_\mathrm{rec}} c_\mathrm{s}(t) dt/a(t)$, the comoving distance that pressure waves can propagate between $t = 0$ and recombination, determines the physical scale of the acoustic peaks in the CMB and the baryon acoustic oscillation (BAO) feature in low redshift matter clustering [22].

- The amplitude of matter fluctuations, conventionally represented by the quantity $\sigma_8(z)$, scales the overall amplitude of growth measures such as weak lensing or redshift-space distortions (discussed in the next section).

Specifically, $\sigma_8(z)$ refers to the rms fluctuation of the matter overdensity $\rho/\bar{\rho}$ in spheres of radius $8 \, h^{-1}\mathrm{Mpc}$, computed from the linear theory matter power spectrum at redshift $z$, and $\sigma_8$ on its own refers to the value at $z = 0$ (just like our convention for $\Omega_\mathrm{m}$).

While discussions of dark energy are frequently phrased in terms of values and errors on quantities like $w_\mathrm{p}$, $w_a$, $\Delta\gamma$, and $\Omega_\mathrm{tot}$, parameter precision is the means to an end, not an end in itself. The underlying goal of empirical studies of cosmic acceleration is to address two physically profound questions:

1. Does acceleration arise from a breakdown of GR on cosmological scales or from a new energy component that exerts repulsive gravity within GR?

2. If acceleration is caused by a new energy component, is its energy density constant in space and time, as expected for a fundamental vacuum energy, or does it show variations that indicate a dynamical field?

Substantial progress towards answering these questions, in particular any definitive rejection of the cosmological constant "null hypothesis," would be a major breakthrough in cosmology and fundamental physics.

## 1.3.    Observational Probes

We briefly summarize the observational probes that play the greatest role in current constraints on dark energy. Further discussion and references can be found in other articles of this *Review*, in particular Secs. 21 (Cosmological Parameters) and 23 (The Cosmic Microwave Background), and in Ref. [12].

*Cosmic Microwave Background Anisotropies:* Although CMB anisotropies provide limited information about dark energy on their own, CMB constraints on the geometry, matter content, and radiation content of the Universe play a critical role in dark energy studies when combined with low redshift probes. In particular, CMB data supply measurements of $\theta_s = r_s/D_{A,c}(z_{rec})$, the angular size of the sound horizon at recombination, from the angular location of the acoustic peaks, measurements of $\Omega_m h^2$ and $\Omega_b h^2$ from the heights of the peaks, and normalization of the amplitude of matter fluctuations at $z_{rec}$ from the amplitude of the CMB fluctuations themselves. Planck data yield a 0.4% determination of $r_s$, which scales as $(\Omega_m h^2)^{-0.25}$ for cosmologies with standard matter and radiation content. The uncertainty in the matter fluctuation amplitude is 3%, dominated by uncertainty in the electron scattering optical depth $\tau$, and it should drop substantially with future analyses of Planck polarization maps. Secondary anisotropies, including the Integrated Sachs-Wolfe effect, the Sunyaev-Zel'dovich (SZ, [23]) effect, and gravitational lensing of primary anisotropies, provide additional information about dark energy by constraining low-redshift structure growth.

*Type Ia Supernovae:* Type Ia supernovae, produced by the thermonuclear explosions of white dwarfs, exhibit 10-15% scatter in peak luminosity after correction for light curve duration (the time to rise and fall) and color (which is a diagnostic of dust extinction). Since the peak luminosity is not known *a priori*, supernova surveys constrain ratios of luminosity distances at different redshifts. If one is comparing a high redshift sample to a local calibrator sample measured with much higher precision (and distances inferred from Hubble's law), then one essentially measures the luminosity distance in $h^{-1}$Mpc, constraining the combination $hD_L(z)$. With distance uncertainties of 5–8% precision per well observed supernova, a sample of $\sim 100$ SNe is sufficient to achieve sub-percent statistical precision. The 1–2% systematic uncertainties in current samples are dominated by uncertainties associated with photometric calibration and dust extinction corrections. Another potential systematic is redshift evolution of the supernova population itself, which can be tested by analyzing subsamples grouped by spectral properties or host galaxy properties to confirm that they yield consistent results.

*Baryon Acoustic Oscillations (BAO):* Pressure waves that propagate in the pre-recombination photo-baryon fluid imprint a characteristic scale in the clustering of matter and galaxies, which appears in the galaxy correlation function as a localized peak at the sound horizon scale $r_s$, or in the power spectrum as a series of oscillations. Since observed galaxy coordinates consist of angles and redshifts, measuring this "standard ruler" scale in a galaxy redshift survey determines the angular diameter distance $D_A(z)$ and the expansion rate $H(z)$, which convert coordinate separations to comoving distances. Errors on the two quantities are correlated, and in existing galaxy surveys the best determined combination is approximately $D_V(z) = [zD_{A,c}^2(z)/H(z)]^{1/3}$. As an approximate rule of thumb, a survey that fully samples structures at redshift $z$ over a comoving volume $V$, and is therefore limited by cosmic variance rather than shot noise, measures $D_{A,c}(z)$ with a fractional error of $0.005(V/10\,\mathrm{Gpc}^3)^{-1/2}$ and $H(z)$ with a fractional error $1.6 - 1.8$ times higher. BAO can also be measured in the Lyman-$\alpha$ forest of intergalactic hydrogen absorption towards background quasars, where the best measured parameter combination is more heavily weighted towards $H(z)$ because of strong redshift-space distortions that enhance clustering along the line of sight. BAO distance measurements complement SN distance measurements by providing absolute rather than relative distances (with precise calibration of $r_s$ from the CMB) and by achieving greater precision at high redshift thanks to the increasing comoving volume available. Theoretical modeling suggests that BAO measurements from even the largest feasible redshift surveys will be limited by statistical rather than systematic uncertainties.

*Weak Gravitational Lensing:* Gravitational light bending by a clustered distribution of matter shears the shapes of higher redshift background galaxies in a spatially coherent manner, producing a correlated pattern of apparent ellipticities. By studying the weak lensing signal for source galaxies binned by photometric redshift (estimated from broad-band colors), one can probe the history of structure growth. For a specified expansion history, the predicted signal scales approximately as $\sigma_8 \Omega_m^\alpha$, with $\alpha \approx 0.3$–$0.5$. The predicted signal also depends on the distance-redshift relation, so weak lensing becomes more powerful in concert with SN or BAO measurements that can pin this relation down independently. The most challenging systematics are shape measurement biases, biases in the distribution of photometric redshifts, and intrinsic alignments of galaxy orientations that could contaminate the lensing-induced signal. Predicting the large-scale weak lensing signal is straightforward in principle, but exploiting small-scale measurements also requires modeling the effects of complex physical processes such as star formation and feedback on the matter power spectrum.

*Clusters of Galaxies:* Like weak lensing, the abundance of massive dark matter halos probes structure growth by constraining $\sigma_8 \Omega_m^\alpha$, where $\alpha \approx 0.3$–$0.5$. These halos can be identified as dense concentrations of galaxies or through the signatures of hot ($10^7$–$10^8$ K) gas in X-ray emission or SZ distortion of the CMB. The critical challenge in cluster cosmology is calibrating the relation $P(M_{\mathrm{halo}}|O)$ between the halo mass as predicted from theory and the observable $O$ used for cluster identification. Measuring the stacked weak lensing signal from clusters has emerged as a promising approach to achieve percent-level accuracy in calibration of the mean relation, which is required for clusters to remain competitive with other growth probes.

*Redshift-Space Distortions (RSD) and the Alcock-Paczynksi (AP) Effect:* Redshift-space distortions of galaxy clustering, induced by peculiar motions, probe structure growth by constraining the parameter combination $f(z)\sigma_8(z)$, where $f(z)$ is the growth rate defined by Eq. (1.3) [25,26]. Uncertainties in theoretical modeling of non-linear gravitational evolution and the non-linear bias between the galaxy and matter distributions currently limit application of the method to large scales (comoving separations $r \gtrsim 10\,h^{-1}\mathrm{Mpc}$ or wavenumbers $k \lesssim 0.2 h\,\mathrm{Mpc}^{-1}$). A second source of anisotropy arises if one adopts the wrong cosmological metric to convert angles and redshifts into comoving separations, a phenomenon known as the Alcock-Paczynksi effect [27]. Demanding isotropy of clustering at redshift $z$ constrains the parameter combination $H(z)D_A(z)$. The main challenge for the AP method is correcting for the anisotropy induced by peculiar velocity RSD.

*Direct Determination of $H_0$:* The value of $H_0$ sets the current value of the critical density $\rho_c = 3H_0^2/8\pi G_N$, and combination with CMB measurements provides a long lever arm for constraining the evolution of dark energy. The challenge in direct $H_0$ measurements is establishing distances to galaxies that are far enough away that their peculiar velocities are small compared to the expansion velocity $v = H_0 d$. This can be done by building a ladder of distance indicators tied to stellar parallax on its lowest rung, or by using gravitational lens time delays or geometrical measurements of maser data to circumvent this ladder.

## 1.4.  Current Constraints on Expansion, Growth, and Dark Energy

The last decade has seen dramatic progress in measurements of the cosmic expansion history and structure growth, leading to much tighter constraints on the parameters of dark energy models. CMB data from the WMAP and Planck satellites and from higher resolution ground-based experiments have provided an exquisitely detailed picture of structure at the recombination epoch and the first CMB-based measures of low redshift structure through lensing and SZ cluster counts. Cosmological supernova samples have increased in size from tens to many hundreds, with continuous coverage from $z = 0$ to $z \approx 1.4$, alongside major improvements in data quality, analysis methods, and detailed understanding of local populations. BAO measurements have advanced from the first detections to 2% precision at multiple redshifts, with increasingly sophisticated methods for testing systematics, fitting models, and evaluating statistical errors. Constraints on low redshift structure from galaxy clusters have become more robust, with improved X-ray and SZ data and weak lensing mass calibrations, and they have been joined by the first precise structure constraints from cosmic shear weak lensing, galaxy-galaxy lensing, and redshift-space distortions. The precision of direct $H_0$ measurements has sharpened from the $\sim 10\%$ error of the HST Key Project [28] to 3–4% in some recent analyses.

As an illustration of current measurements of the cosmic expansion history, Figure 1.1 compares distance-redshift measurements from SN and BAO data to the predictions for a flat universe with a cosmological constant. SN cosmology relies on compilation analyses that try to bring data from different surveys probing distinct redshift ranges to a common scale. The most influential current compilations are SNLS3 [29], which combines data from the 3-year Supernova Legacy Survey sample and the 1st-year SDSS-II Supernova Survey sample with local calibrators and high-redshift SNe from HST surveys, and Union2.1 [30], which has a broader selection of data, including some but not all of the sources in SNLS3. Here we have used binned distance measurements from Union2.1, but we caution that the different sample selections and analysis methodologies lead to systematic differences comparable to the statistical uncertainties, and it is not obvious which compilation, if either, should be preferred. Because the peak luminosity of a fiducial SN Ia is an unknown free parameter, the SN distance measurements could all be shifted up and down by a constant multiplicative factor; cosmological information resides in the relative distances as a function of redshift. The four BAO data points are taken from analyses of the 6dFGS survey [31], SDSS-II [32], BOSS [33], and WiggleZ [34]. For the BAO measurements we have adopted the sound horizon scale $r_s = 147.49$ Mpc from Planck CMB data, whose 0.4% uncertainty is small compared to the current BAO measurement errors [36]. We have converted both SN luminosity distances and BAO $D_V$ distances to an equivalent comoving angular diameter distance.

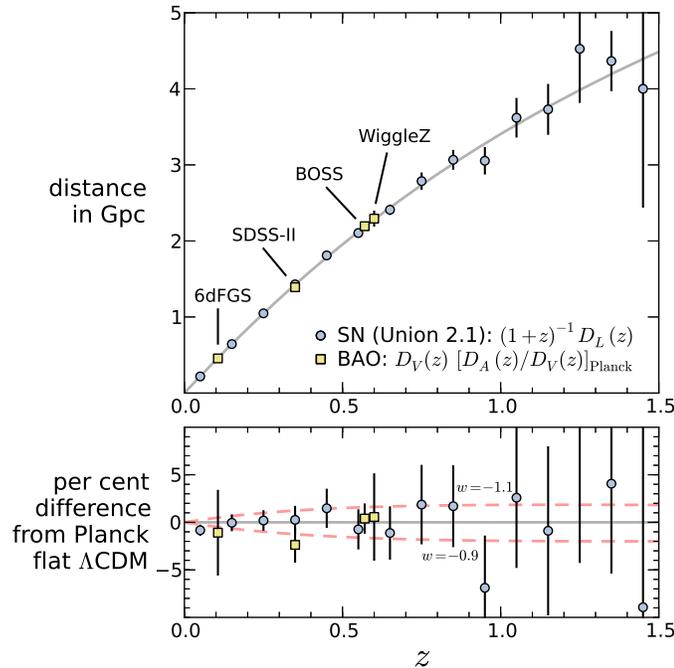

**Figure 1.1:** The distance-redshift relation measured from Type Ia SNe and BAO compared to the predictions (gray curve) of a flat ΛCDM model with the best-fit parameters inferred from Planck+WP CMB data. Circles show binned luminosity distances from the Union2.1 SN sample, multiplied by $(1 + z)^{-1}$ to convert to comoving angular diameter distance. Squares show BAO distance measurements, converted to $D_{A,c}(z)$ for the Planck+WP cosmology and sound horizon, from the references given in the text. The lower panel plots residuals from the Planck+WP ΛCDM prediction, with dashed curves that show the effect of changing $w$ by $\pm 0.1$ while all other parameters are held fixed. Note that the SN data points can be shifted up or down by a constant factor to account for freedom in the peak luminosity, while the BAO points are calibrated to 0.4% precision by the sound horizon scale computed from Planck+WP data.

The plotted cosmological model has $\Omega_m = 0.315$ and $h = 0.673$, the best-fit values [37] from Planck+WP CMB data assuming $w = -1$ and $\Omega_{tot} = 1$. Specifically, here and below we use parameter values and MCMC chains from the "Planck + WP" analysis of [38], which combines the Planck temperature power spectrum with low multipole polarization measurements from WMAP [39]. In contrast to the Cosmological Parameters article of this *Review*, we do not use the CMB data set that includes higher resolution ground-based results because the corresponding chains are not available for all of the cases we wish to examine, but differences in cases where they are available are small. The SN, BAO, and CMB data sets, probing a wide range of redshifts with radically different techniques, are mutually consistent with the predictions of a flat ΛCDM cosmology. We have not included the $z = 2.5$ BAO measurement from the BOSS Lyman-$\alpha$ forest [24] on this plot, but it is also consistent with this fiducial model. Other curves in the lower panel of Figure 1.1 show the effect of changing $w$ by $\pm 0.1$ with all other parameters held fixed. However, such a single-parameter comparison does not capture the impact of parameter degeneracies or the ability of complementary data sets to break them, and if one instead forces a match to CMB data by changing $h$ and $\Omega_m$ when changing $w$ then the predicted BAO distances diverge at $z = 0$ rather than converging there.

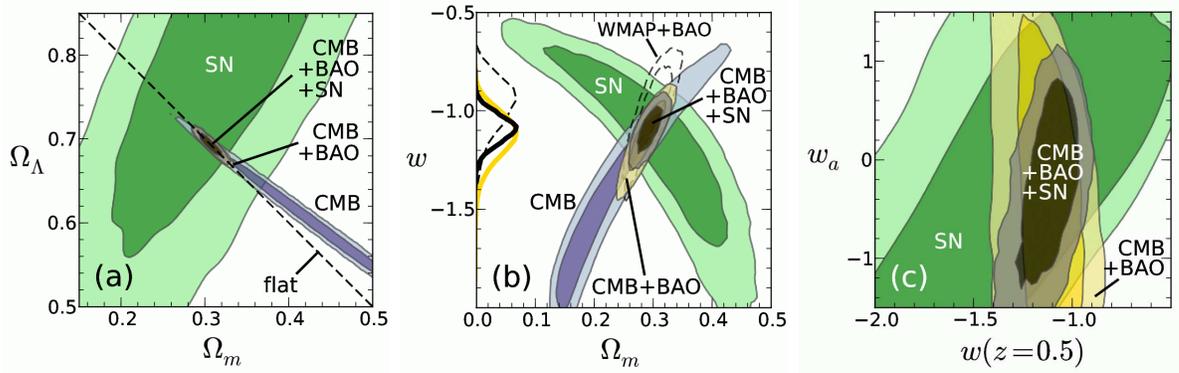

**Figure 1.2:** Constraints on the present matter fraction $\Omega_m$ and dark energy model parameters. Dark and light shaded regions indicate 68.3% and 95.4% confidence levels, respectively. "CMB" is Planck+WP, "BAO" is the combination of SDSS-II, BOSS, and 6dFGS, and "SN" is Union2. (a) The present dark energy fraction $\Omega_\Lambda$ vs. $\Omega_m$, assuming a ΛCDM model. CMB data, especially when combined with BAO constraints, strongly favor a flat universe (diagonal dashed line). (b) The dark energy equation of state $w$ vs. $\Omega_m$, assuming a constant value of $w$. The dashed contours show the 68.3% and 95.4% CL regions for the combination of WMAP9 and BAO data. Curves on the left vertical axis show the probability distributions for $w$ (normalized arbitrarily), after marginalizing over $\Omega_m$, for the CMB+BAO and CMB+BAO+SN combinations (yellow and black, respectively), using Planck+WP CMB data, and for the WMAP9+BAO combination (dashed black). (c) Constraints on the two parameters of the dark energy model with a time-dependent equation of state given by Eq. (1.4): $w(z = 0.5)$ and $w_a = -dw/da$.

Figure 1.2a plots joint constraints on $\Omega_m$ and $\Omega_\Lambda$ in a ΛCDM cosmological model, assuming $w = -1$ but not requiring spatial flatness. The SN constraints are computed from the Union2 sample, and the CMB, CMB+BAO, and CMB+BAO+SN constraints are taken from MCMC chains provided by the Planck Collaboration [38]. We do not examine BAO constraints separately from CMB, because the constraining power of BAO relies heavily on the CMB calibration of $r_s$. The SN data or CMB data on their own are sufficient to reject an $\Omega_\Lambda = 0$ universe, but individually they allow a wide range of $\Omega_m$ and significant non-zero curvature. The CMB+BAO combination zeroes in on a tightly constrained region with $\Omega_m = 0.309 \pm 0.011$ and $\Omega_{tot} = 1.000 \pm 0.0033$. Combining SN with CMB would lead to a consistent constraint with around 3–4× larger errors.

Adding the SN data to the CMB+BAO combination makes only a small difference to the constraints in this restricted model space.

Figure 1.2b plots constraints in the $\Omega_\mathrm{m} - w$ space, where we now consider models with constant $w(z)$ and (in contrast to panel a) assume spatial flatness. CMB data alone allow a wide range of $w$, but combination with BAO narrows the allowed range sharply. The preferred region is consistent with the orthogonal SN constraint, and the combination of the three data sets yields smaller uncertainties. The black curve on the left axis shows the posterior p.d.f. for $w$ after marginalizing (with a flat prior) over $\Omega_\mathrm{m}$; we find $w = -1.10 \pm 0.08$ at 68.3% CL and $-1.10 \pm 0.15$ at 95.4% CL. The dashed contours and dashed marginal curve show the impact of substituting WMAP9 data for Planck+WP in the CMB+BAO combination. The two constraints are compatible, but the shift from WMAP to Planck+WP has reduced the uncertainty in $w$ and pulled the best-fit value lower.

Figure 1.2c considers a model space with time varying $w$, evolving according to the linear parameterization $w(a) = w_0 + w_a(1 - a)$, again assuming flat space. Instead of $w_0$ we show constraints on $w(z = 0.5)$, approximately the pivot redshift where $w$ is best determined and covariance with $w_a$ is minimized. This plot shows that even the combination of current CMB, BAO, and SN data places only weak constraints on time evolution of the equation of state, still allowing order unity changes in $w$ between $z = 1$ and $z = 0$ ($\Delta a = 0.5$). The value of $w(z = 0.5)$, on the other hand, is reasonably well constrained, with errors only slightly larger than those for the constant-$w$ model of panel b. Errors on $w_0 = w(z = 0.5) - 0.333 w_a$ are much larger and are strongly correlated with the $w_a$ errors.

While the CMB, BAO, and SN data sets considered here are mutually consistent with a flat $\Lambda$CDM model, tensions arise when other cosmological measurements enter the mix. Blue and yellow contours in Figure 1.3a show CMB and CMB+BAO constraints in the $\Omega_\mathrm{m} - H_0$ plane, assuming $w = -1$ and $\Omega_\mathrm{tot} = 1$. Red horizontal bars represent the direct estimate $H_0 = 73.8 \pm 2.4 \, \mathrm{km \, s^{-1} \, Mpc^{-1}}$ from Ref. [40], who use SN Ia distances to galaxies in the Hubble flow with the Ia luminosity scale calibrated by HST observations of Cepheids in nearby SN host galaxies. Another recent estimate by Ref. [41], which employs $3.6 \, \mu m$ Cepheid observations to recalibrate the HST Key Project distance ladder and reduce its uncertainties, yields a similar central value and estimated error, $H_0 = 74.3 \pm 2.1 \, \mathrm{km \, s^{-1} \, Mpc^{-1}}$. Figure 1.3a indicates a roughly $2\sigma$ tension between these direct measurements and the CMB+BAO predictions. The tension was already present with WMAP CMB data, as shown in Figure 1.3b, but it has become stiffer with Planck+WP, because of smaller CMB+BAO errors and a shift of central values to slightly higher $\Omega_\mathrm{m}$ and lower $H_0$. In models with free, constant $w$ (still assuming $\Omega_\mathrm{tot} = 1$), the tension can be lifted by going to $w < -1$ and lower $\Omega_\mathrm{m}$, as illustrated in Figure 1.3c. CMB data determine $\Omega_\mathrm{m} h^2$ with high precision from the heights of the acoustic peaks, essentially independent of $w$. Within the flat $\Lambda$CDM framework, the well determined distance to the last scattering surface pins down a specific combination of $(\Omega_\mathrm{m}, h)$, but with free $w$ one can obtain the same distance from other combinations along the $\Omega_\mathrm{m} h^2$ degeneracy axis.

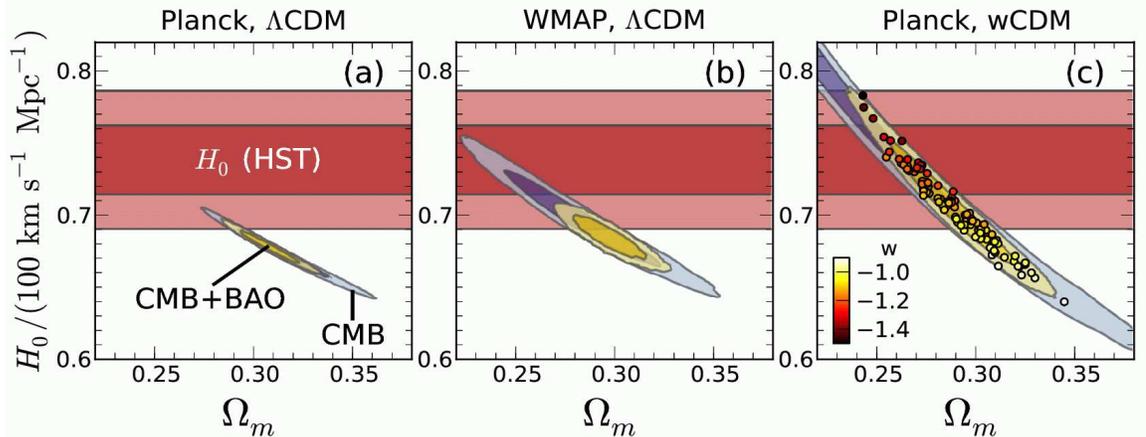

**Figure 1.3:** Constraints on the present matter fraction $\Omega_m$ and the Hubble constant $H_0$ from various combinations of data, assuming flat $\Lambda$CDM (left and middle panels) or a constant dark energy equation of state $w$ (right panel). Dark and light shaded regions indicate 68.3% and 95.4% confidence levels, respectively. The right panel also shows 100 Monte Carlo samples from the CMB+BAO constraints with the value of $w$ indicated by the colors of the dots. "CMB" is Planck+WP in the outer panels and WMAP9 in the middle panel, "BAO" is the combination of SDSS-II, BOSS, and 6dFGS, and "$H_0$ (HST)" is the HST constraint from [40].

One should not immediately conclude from Figure 1.3 that $w \neq -1$, but this comparison highlights the importance of fully understanding (and reducing) systematic uncertainties in direct $H_0$ measurements. If errors were reduced and the central value remained close to that plotted in Figure 1.3, then the implications would be striking. Other recent $H_0$ determinations exhibit less tension with CMB+BAO, because of lower central values and/or larger errors [42,43], including the values of $H_0 = 69 \pm 7 \,\mathrm{km\,s^{-1}\,Mpc^{-1}}$ and $68 \pm 9 \,\mathrm{km\,s^{-1}\,Mpc^{-1}}$ from Refs. [44,45], who circumvent the traditional distance ladder by using maser distances to galaxies in the Hubble flow. Gravitational lens time delays offer another alternative to the traditional distance ladder, and their precision could become competitive over the next few years, with increasing sample sizes and better constrained lens models.

The amplitude of CMB anisotropies is proportional to the amplitude of density fluctuations present at recombination, and by assuming GR and a specified dark energy model one can extrapolate the growth of structure forward to the present day to predict $\sigma_8$. As discussed in Sec. 1.3 probes of low redshift structure typically constrain the combination $\sigma_8 \Omega_m^\alpha$ with $\alpha \approx 0.3$–$0.5$. Figure 1.4 displays constraints in the $\sigma_8 - \Omega_m$ plane from CMB+BAO data and from weak lensing and cluster surveys [46]. Planck data themselves reveal a CMB lensing signature that constrains low redshift matter clustering and suggests a fluctuation amplitude somewhat lower than the extrapolated value for flat $\Lambda$CDM. However, including the CMB lensing signal only slightly alters the Planck+WP confidence interval for $\Lambda$CDM (purple vs. yellow contours in Fig. 1.4a). Allowing free $w$ (gray contours) expands this interval, primarily in the direction of lower $\Omega_m$ and higher $\sigma_8$ (with $w < -1$).

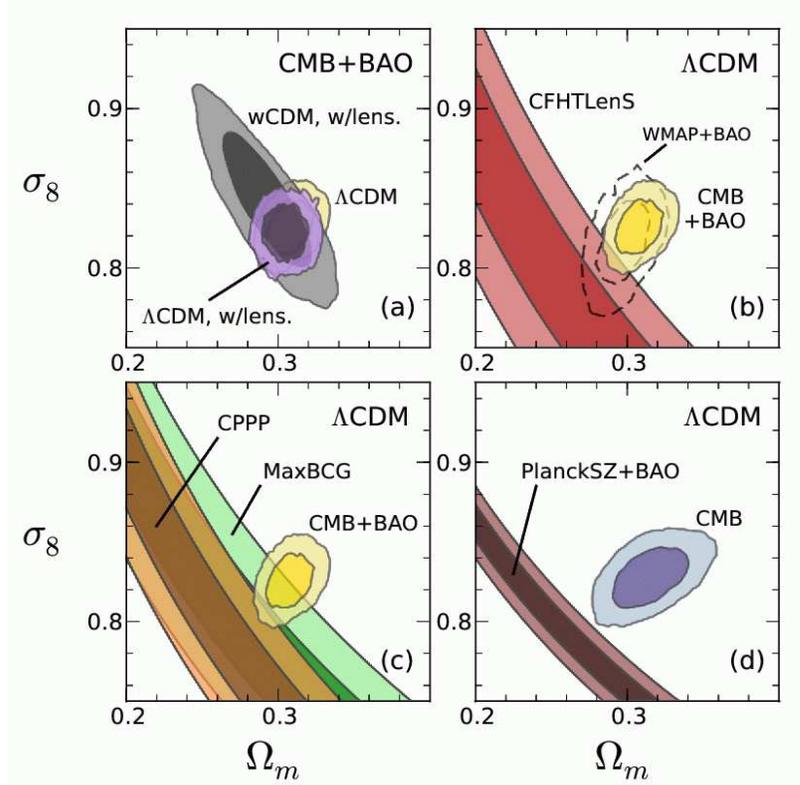

**Figure 1.4:** Constraints on the present matter fraction $\Omega_{\mathrm{m}}$ and the present matter fluctuation amplitude $\sigma_8$. Dark and light shaded regions indicate 68.3% and 95.4% confidence levels, respectively. The upper left panel compares CMB+BAO constraints (using the same data sets as in Fig. 1.2) for $\Lambda$CDM with and without CMB lensing, and for a constant $w$ model (including CMB lensing). The other three panels compare flat $\Lambda$CDM constraints between various dark energy probes, including weak lensing (upper right panel) and clusters (lower panels).

The red contours in Figure 1.4b plot the constraint $\sigma_8(\Omega_{\mathrm{m}}/0.27)^{0.46} = 0.774^{+0.032}_{-0.041}$ inferred from tomographic cosmic shear measurements in the CFHTLens survey [47]. An independent analysis of galaxy-galaxy lensing and galaxy clustering in the SDSS yields a similar result [48], $\sigma_8(\Omega_{\mathrm{m}}/0.27)^{0.57} = 0.77 \pm 0.05$. Note that $\sigma_8$ and $\Omega_{\mathrm{m}}$ refer to $z = 0$ values; the weak lensing samples and the cluster samples discussed below are not at zero redshift, but the values of $\sigma_8$ are effectively extrapolated to $z = 0$ for a fiducial cosmology. (Within current parameter bounds, the uncertainty in extrapolating growth from $z = 0.5$ to $z = 0$ is 1–2%, small compared to the observational uncertainties.) There is approximately $2\sigma$ tension between the $\sigma_8 - \Omega_{\mathrm{m}}$ combination predicted by Planck+WP CMB+BAO for $\Lambda$CDM and the lower value implied by the weak lensing measurements. This tension was weaker for WMAP+BAO data (dotted contour) because of the larger error and slightly lower best-fit parameter values.

Additional contours in Figures 1.4c and d show $\sigma_8 - \Omega_{\mathrm{m}}$ constraints inferred from three representative cluster analyses [49]: $\sigma_8(\Omega_{\mathrm{m}}/0.27)^{0.47} = 0.784 \pm 0.027$ (CPPP), $\sigma_8(\Omega_{\mathrm{m}}/0.27)^{0.41} = 0.806 \pm 0.032$ (MaxBCG), and $\sigma_8(\Omega_{\mathrm{m}}/0.27)^{0.32} = 0.782 \pm 0.010$ (PlanckSZ). The basic mass calibration comes from X-ray data in CPPP, from weak lensing data in MaxBCG, and from SZ data in PlanckSZ. Because the PlanckSZ constraint itself incorporates BAO data, we have replaced the CMB+BAO contour with a CMB-only contour in panel d. The $\sigma_8 \Omega_m^\alpha$ constraints from recent cluster analyses are not in perfect agreement, and the examples shown here are far from exhaustive. Nonetheless, on balance the cluster analyses, like the weak lensing analyses, favor lower $\sigma_8 \Omega_m^\alpha$ than the value extrapolated forward from Planck+WP assuming flat $\Lambda$CDM. Redshift-space distortion analyses also tend to favor lower $\sigma_8 \Omega_m^\alpha$, though statistical errors are still fairly large. For example, [50] find $f(z)\sigma_8(z) = 0.415 \pm 0.034$ from SDSS-III BOSS galaxies at $z = 0.57$, while the best-fit Planck+WP+BAO flat $\Lambda$CDM model predicts $f(z)\sigma_8(z) = 0.478 \pm 0.008$ at this redshift. With somewhat more aggressive modeling assumptions, [51] infer $f(z)\sigma_8(z)$ from the WiggleZ survey at $z = 0.22$, 0.41, 0.60, and 0.78, with $\approx 10\%$ errors in the three highest redshift bins (and 17% at $z = 0.22$), finding excellent agreement with a flat $\Lambda$CDM model that has $\Omega_m = 0.27$ and $\sigma_8 = 0.8$, and thus with the structure measurements plotted in Figure 1.4.

Going from $\Lambda$CDM to $w$CDM does not readily resolve this tension, because the CMB degeneracy direction with free $w$ is roughly parallel to the $\sigma_8 \Omega_m^\alpha$ tracks from low redshift structure (though the tracks themselves could shift or widen for $w \neq -1$). Each of the low redshift probes has significant systematic uncertainties that may not be fully represented in the quoted observational errors, and the tensions are only about $2\sigma$ in the first place, so they may be resolved by larger samples, better data, and better modeling. However, it is notable that all of the discrepancies are in the same direction. On the CMB side, the tensions would be reduced if the value of $\Omega_m$ or the optical depth $\tau$ (and thus the predicted $\sigma_8$) has been systematically overestimated. The most exciting but speculative possibility is that these tensions reflect a deviation from GR-predicted structure growth, pointing towards a gravitational explanation of cosmic acceleration. Other possible physical resolutions could come from dark energy models with significant time evolution, from a massive neutrino component that suppresses low redshift structure growth, or from decaying dark matter that reduces $\Omega_m$ at low $z$.

Table 1.1 summarizes key results from Figures 1.2–1.4, with marginalized constraints on $\Omega_m$, $\Omega_{tot}$, $w$, $h$, and $\sigma_8(\Omega_m/0.27)^{0.4}$ for the Planck+WP+BAO, Planck+WP+BAO+SN, and WMAP9+BAO combinations. We list 68.3% errors, and also 95.4% errors for WMAP9+BAO constraints on $w$CDM; in all other cases, the 95.4% errors are very close to double the 68.3% errors. For $\Lambda$CDM the Planck+WP combinations give $\Omega_{tot} = 1.000$ with an error of 0.3% and they predict, approximately, $h = 0.68 \pm 0.01$ and $\sigma_8(\Omega_m/0.27)^{0.4} = 0.87 \pm 0.02$. Note that the $\Omega_m$ and $h$ constraints are not identical to those in Table 21.1 of the Cosmological Parameters article of this *Review* because those values assume spatial flatness. For $w$CDM, where flatness *is* assumed, the Planck+WP+BAO+SN combination yields $w = -1.10^{+0.08}_{-0.07}$, consistent with a cosmological constant at $1.2\sigma$. With free $w$ the best-fit $h$ increases and its error roughly doubles, but the error in $\sigma_8(\Omega_m/0.27)^{0.4}$ grows only slightly, and its best-fit value moves a bit further away from the lower amplitudes suggested by measurements of low redshift structure.

**Table 1.1:** Constraints on selected parameters from various combinations of CMB, BAO, and SN data, given as mean values $\pm$ 68.3% CL limits (and $\pm$ 95.4% CL limits for WMAP9-$w$CDM). "Planck+WP" combines the Planck temperature power spectrum with WMAP large scale polarization. "BAO" combines the measurements of SDSS-II, BOSS, and 6dFGS. "SN" refers to the Union2.1 compilation. The upper (lower) half of the table assumes a $\Lambda$CDM (flat $w$CDM) cosmological model.

| | Data combination | | |
|---|---|---|---|
| Parameter | Planck+WP+BAO | Planck+WP+BAO+SN | WMAP9+BAO |
| **$\Lambda$CDM** | | | |
| $\Omega_{\rm m}$ | $0.309^{+0.010}_{-0.011}$ | $0.307^{+0.011}_{-0.010}$ | $0.295^{+0.012}_{-0.012}$ |
| $\Omega_{\rm tot}$ | $1.000^{+0.0033}_{-0.0033}$ | $1.000^{+0.0032}_{-0.0033}$ | $1.003^{+0.004}_{-0.004}$ |
| $h$ | $0.678^{+0.011}_{-0.010}$ | $0.679^{+0.010}_{-0.011}$ | $0.681^{+0.011}_{-0.011}$ |
| $\sigma_8(\Omega_{\rm m}/0.27)^{0.4}$ | $0.871^{+0.020}_{-0.021}$ | $0.869^{+0.020}_{-0.021}$ | $0.836^{+0.033}_{-0.033}$ |
| **$w$CDM (flat)** | | | |
| $\Omega_{\rm m}$ | $0.287^{+0.021}_{-0.021}$ | $0.294^{+0.014}_{-0.014}$ | $0.299^{+0.022}_{-0.019}\left(^{+0.045}_{-0.042}\right)$ |
| $w$ | $-1.13^{+0.13}_{-0.11}$ | $-1.10^{+0.08}_{-0.07}$ | $-0.98^{+0.16}_{-0.12}\left(^{+0.33}_{-0.29}\right)$ |
| $h$ | $0.708^{+0.026}_{-0.030}$ | $0.699^{+0.017}_{-0.018}$ | $0.681^{+0.025}_{-0.032}\left(^{+0.060}_{-0.066}\right)$ |
| $\sigma_8(\Omega_{\rm m}/0.27)^{0.4}$ | $0.888^{+0.025}_{-0.025}$ | $0.885^{+0.023}_{-0.023}$ | $0.84^{+0.05}_{-0.05}\left(^{+0.09}_{-0.09}\right)$ |

## 1.5.   Summary and Outlook

The preceding figures and table focus on model parameter constraints, but as a description of the observational situation it is most useful to characterize the precision, redshift range, and systematic uncertainties of the basic expansion and growth measurements. At present, supernova surveys constrain distance ratios at the 1–2% level in redshift bins of width $\Delta z = 0.1$ over the range $0 < z < 0.6$, with larger but still interesting error bars out to $z \approx 1.2$. These measurements are currently limited by systematics tied to photometric calibration, extinction, and reddening, and possible evolution of the SN population. BAO surveys have measured the absolute distance scale (calibrated to the sound horizon $r_{\rm s}$) to 4.5% at $z = 0.11$, 2% at $z = 0.35$ and $z = 0.57$, 6% at $z = 0.73$, and 3% at $z = 2.5$. Multiple studies have used clusters of galaxies or weak lensing cosmic shear or galaxy-galaxy lensing to measure a parameter combination $\sigma_8\Omega_{\rm m}^\alpha$ with $\alpha \approx 0.3$–0.5. The estimated errors of these studies, including both statistical contributions and identified systematic uncertainties, are about 5%. RSD measurements constrain the combination $f(z)\sigma_8(z)$, with recent determinations spanning the redshift range $0 < z < 0.9$ with typical estimated errors of about 10%. These errors are dominated by statistics, but shrinking them further will require improvements in modeling non-linear effects on small scales. Direct distance-ladder estimates of $H_0$ now span a small range (using overlapping data but distinct treatments of key steps), with individual studies quoting uncertainties of 3–5%, with similar statistical and systematic contributions. Planck data and higher resolution ground-based experiments now measure CMB anisotropy with exquisite precision.

A flat $\Lambda$CDM model with standard radiation and neutrino content can fit the CMB data and the BAO and SN distance measurements to within their estimated uncertainties. However the Planck+WP+BAO parameters for this model are in approximately $2\sigma$ tension with some of the direct $H_0$ measurements and most of the cluster and weak lensing analyses, disagreeing by about 10% in each case. Similar tensions are present when using WMAP data in place of Planck+WP data, but they are less evident because the WMAP errors are larger and the best-fit $\Omega_{\rm m}$ value is lower. Moving from $\Lambda$CDM to $w$CDM can relieve the tension with $H_0$, but only by going to $w < -1$ (which would be more physically startling than $w > -1$), and this change on its own does not produce better agreement with the structure growth data. It is not clear whether current tensions should be taken as a sign of new physics or as a sign that at least some of the experiments are underestimating their systematic uncertainties. Factor-of-two reductions in error bars, if convincing, could lead to exciting physical implications, or to a resolution of the existing mild discrepancies. Moving forward, the community will have to balance the requirement of strong evidence for interesting claims (such as $w \neq -1$ or deviations from GR) against the danger of confirmation bias, *i.e.*, discounting observations or error estimates when they do not overlap simple theoretical expectations.

There are many ongoing projects that should lead to improvement in observational constraints in the near-term and over the next two decades [52]. Final analyses of Planck temperature and polarization maps will significantly tighten the CMB constraints, including an important reduction of the uncertainty in the matter fluctuation amplitude that will sharpen tests based on structure growth. Final data from the SDSS-III BOSS survey, finishing in 2014, will reduce BAO errors by a factor of two at $z = 0.3$, 0.6, and 2.5. Its SDSS-IV successor eBOSS will yield the first BAO measurements in the redshift range $1 < z < 2$ and improved precision at lower and higher redshifts. The HETDEX project will measure BAO with Lyman-$\alpha$ emission line galaxies at $z = 2$–3. The same galaxy surveys carried out for BAO also provide data for RSD measurements of structure growth and AP measurements of cosmic geometry, and with improved theoretical modeling there is potential for large precision gains over current constraints from these methods. The Dark Energy Survey (DES), which started operations in August 2013 and will run through 2018, will provide a sample of several thousand Type Ia SNe, enabling smaller statistical errors and division of the sample into subsets for cross-checking evolutionary effects and other systematics. DES imaging will be similar in depth but 50 times larger in area than CFHTLens, providing a much more powerful weak lensing data set and weak lensing mass calibration of enormous samples of galaxy clusters (tens of thousands).

Weak lensing surveys from the newly commissioned Hyper Suprime-Cam on the Subaru telescope will be smaller in area but deeper, with a comparable number of lensed galaxies. Reducing weak lensing systematics below the small statistical errors of these samples will be a major challenge, but one with a large payoff in precision measurements of structure growth. Uncertainties in direct determinations of $H_0$ should be reduced by further observations with HST and, in the longer run, by Cepheid parallaxes from the GAIA mission, by the ability of the James Webb Space Telescope to discover Cepheids in more distant SN Ia calibrator galaxies, and by independent estimates from larger samples of maser galaxies and gravitational lensing time delays.

A still more ambitious period begins late in this decade and continues through the 2020s, with experiments that include the Dark Energy Spectroscopic Instrument (DESI), the Subaru Prime Focus Spectrograph (PFS), the Large Synoptic Survey Telescope (LSST), and the space missions Euclid and WFIRST (Wide Field Infrared Survey Telescope). DESI and PFS both aim for major improvements in the precision of BAO, RSD, and other measurements of galaxy clustering in the redshift range $0.8 < z < 2$, where large comoving volume allows much smaller cosmic variance errors than low redshift surveys like BOSS. LSST will be the ultimate ground-based optical weak lensing experiment, measuring several billion galaxy shapes over $20,000 \deg^2$ of the southern hemisphere sky, and it will detect and monitor many thousands of SNe per year. Euclid and WFIRST also have weak lensing as a primary science goal, taking advantage of the high angular resolution and extremely stable image quality achievable from space. Both missions plan large spectroscopic galaxy surveys, which will provide better sampling at high redshifts than DESI or PFS because of the lower infrared sky background above the atmosphere. WFIRST is also designed to carry out what should be the ultimate supernova cosmology experiment, with deep, high resolution, near-IR observations and the stable calibration achievable with a space platform.

Performance forecasts necessarily become more uncertain the further ahead we look, but collectively these experiments are likely to achieve 1–2 order of magnitude improvements over the precision of current expansion and growth measurements, while simultaneously extending their redshift range, improving control of systematics, and enabling much tighter cross-checks of results from entirely independent methods. The critical clue to the origin of cosmic acceleration could also come from a surprising direction, such as laboratory or solar system tests that challenge GR, time variation of fundamental "constants," or anomalous behavior of gravity in some astronomical environments. Experimental advances along these multiple axes could confirm today's relatively simple, but frustratingly incomplete, "standard model" of cosmology, or they could force yet another radical revision in our understanding of energy, or gravity, or the spacetime structure of the Universe.